\documentclass{elsart}
\usepackage[square]{natbib}
\usepackage{graphicx}
\journal{Ocean Dynamics}

\begin{document}
\begin{frontmatter}
\title{Local dynamical equivalence of certain food webs}
\author{Thilo~Gross\corauthref{cor}}
\corauth[cor]{ Corresponding author.\\ 
              {\it Email:} thilo.gross@physics.org }
\address{Max-Planck Institute for Physics of Complex Systems, N\"othnitzer Stra\ss e 38, 01187 Dresden, Germany} 
\author{Ulrike Feudel}
\address{Institut f\"ur Chemie und Biologie des Meeres, Carl von Ossietzky Universit\"at,PF 2503, 26111 Oldenburg, Germany.}
\date{\today}

\begin{abstract}
An important challenge in theoretical ecology is to find good-coarse grained representations of complex 
food webs. Here we use the approach of generalized modeling to show that it may be possible 
to formulate a coarse-graining algorithm that conserves the local dynamics of the model exactly. We show 
examples of food webs with a different number of species that have exactly identical local bifurcation 
diagrams. Based on these observations, we formulate a conjecture governing which populations of complex food webs can be grouped together into a single variable without changing the local dynamics. As an illustration we use this conjecture to show that chaotic regions generically exist in the parameter space of a class of food webs with more than three trophic levels. While our conjecture is at present only applicable to relatively
special cases we believe that its applicability could be greatly extended if a more sophisticated mapping 
of parameters were used in the model reduction. 
\end{abstract}
\end{frontmatter}

\section{Introduction}
\label{secIntro} 
Over the recent decades there has been significant progress in ecological modeling. However, 
this progress manifests itself mostly in the description of small systems containing only few 
species. By contrast, to predict the dynamics of large ecosystems remains an important open challenge. In the context of marine ecosystems it has often been pointed out that several obstacles 
have to be overcome in order to model even the planktonic food web, which forms their backbone, realistically .

One obstacle that is encountered in plankton modeling is the sheer diversity of marine life. The 
traditional approach of ecological modeling is to describe every single population by at least one 
differential equation. If we applied this approach to a marine food web, we would obtain a huge 
system, which would be prohibitively difficult to study both analytically and numerically. 
A promising alternative is therefore to model the food web not on the level of the population, but on 
a coarse-grained level at which every variable describes several similar populations. 

Coarse-grained descriptions have a long tradition in plankton modeling. The earliest models, the N-P-Z
food chains, consisted only of three equations describing, nutrients N, phytoplankton P, and zooplankton Z (see e.g. \cite{Steele:Predation}). It is clear that models at this level of abstraction cannot capture the dynamics of the planktonic food web in any detail. However, they have the advantage that the categories N, P, and Z can be clearly defined. Many current models use more refined categories such as guilds and functional groups. In this case it can be unclear if a given population should belong to one category, or the other, or should be a category on its own \cite{Anderson:Running}. The traditional solution is to group populations that share certain biological characteristics such as feeding behavior, metabolism, or activity cycle. Some newer approaches use the tools of graph theory to identify \emph{cliques} or \emph{communities} that hold a similar topological position in the food web \cite{Newman:Clique}. The former approach ensures that populations that are grouped together in one category, fill a similar role in the ecosystem; the latter, that they hold a similar topological position. However, does either of the one imply that the grouped species have a similar impact on the dynamics?

In the formulation of a coarse-grained food web model the goal should be to group species into categories such that the dynamics of the system is not changed by the coarse-graining, at least, not qualitatively. However, to check whether this condition is satisfied does not seem to be feasible as it requires knowledge of the dynamics of the original system which we set out to determine in the first place. Nevertheless, if an approach is available to extract at least certain dynamical properties from the full model we can check whether these dynamical properties are conserved in the coarse-grained model. 

In this paper paper we apply the approach of generalized modeling to explore the effect of coarse-graining on classes of simple food webs. Generalized modeling can extract certain dynamical properties, i.e., the local dynamics around steady states, very efficiently. One advantage of generalized models is that a single generalized model does not describe a single food web, but the whole class of food webs which share a similar topology. This enables us to check whether a given coarse-graining step qualitatively preserves the local dynamics in a given class of food webs. Based on these results we conjecture that certain food webs can be coarse-grained to food chains with the same number of trophic levels. This allows us to generalize results on food chains to a class of food webs. In particular we present evidence that parameter regions in which chaotic dynamics take place, should exist in many food-chains with more than three trophic levels.           

The paper is organized as follows: We start by considering some simple examples in Sec.~\ref{secIllustrative}.
The insights gained in this section lead to the formulation of some conjecture which is stated 
in Sec.~\ref{secGeneralized}. This conjecture is then applied to generalize previous results on chaotic parameter regions in food chains to a class of food webs. Finally, we summarize and discuss our findings in Sec.~\ref{secConclusions}.
 
\section{Two illustrative examples}
Let us start by using generalized models to study two simple examples. The first example is a generalized predator-prey system with one predator. We use it to give a brief introduction to the approach of generalized modeling. Thereafter we study a system of two populations of predators competing for one prey population. In this second example we discuss the generalized modeling only briefly, but focus instead on the results of the modeling process: In certain cases the local dynamics of the system is identical to the one observed in the predator-prey system. So, in order to capture the dynamics of the system it is sufficient to describe the two populations of predators by a single variable.
 
\subsection{The generalized predator-prey system \label{secIllustrative}}
Let us study the dynamics of a generalized predator prey system. The system consists of a prey population $X$ and a predator population $Y$. In the absence of the predator, the prey population grows at the rate $S(X)$. If the predator is introduced it consumes prey at the rate $G(X,Y)$ and loses individuals because of natural mortality at the rate $M(Y)$. In the following we will not restrict $S$, $G$ and $M$ to specific functional forms, but study the generalized model given by
\begin{equation}
\label{eqPredPrey}
\begin{array}{r c l}
\dot{X}&=&S(X)-G(X,Y)\\
\dot{Y}&=&G(X,Y)-M(Y)
\end{array}
\end{equation}
where $S$, $G$, and $M$ are arbitrary positive functions. The equations seem to suggest that all the biomass that is consumed by the predator is actually converted to predator biomass. Note however, that any non-vanishing conversion efficiency that may exist in a given system can always be set to unity by choosing the units of predator or prey biomass appropriately. 

In conventional modeling the first step of model analysis is often to compute the steady states of the system under consideration. In the generalized model this is impossible with the desired degree of generality. But, there is in fact no need to compute steady states: The generalized model describes a whole class of systems, which covers the state space densely with steady states. In other words, for every steady $(X^*,Y^*)$ we observe in nature there is at least one specific model of the form of Eq.~(\ref{eqPredPrey}) that has a steady state exactly at this point. In fact there is even a whole family of infinitely many specific models that have a steady state at $(X^*,Y^*)$. 

Let us assume we are interested in an arbitrary steady state $(X^*,Y^*)$. We start our investigations by normalizing the state variables and the biomass fluxes to unity. For this purpose we define new variables
\begin{equation}
x=\frac{X}{X^*},\; y=\frac{Y}{Y^*}
\end{equation}         
and functions 
\begin{equation}
s(x)=\frac{S(xX^*)}{S^*},\; g(x,y)=\frac{G(xX^*,yY^*)}{G^*},\; m(y)=\frac{M(yY^*)}{M^*}
\end{equation}
where $S^*$, $G^*$, and $M^*$ denote the fluxes in the steady state $S(X^*)$, $G(X^*,Y^*)$, and $M(Y^*)$.

Substituting the definitions into Eq.~(\ref{eqPredPrey}) yields
\begin{equation}
\label{eqPredPreyNorm}
\begin{array}{r c l}
\displaystyle \dot{x}&=&\frac{S^*}{X^*}s(x)-\frac{G^*}{X^*}g(x,y)\\
\displaystyle \dot{y}&=&\frac{G^*}{Y^*}g(x,y)-\frac{M^*}{Y^*}m(y)
\end{array}
\end{equation} 
The structure of this system of equations is still the same as in Eq.~(\ref{eqPredPrey}), but some unsightly pre-factors have appeared. To get rid of these we define the constants 
\begin{equation}
\label{eqParamDef}
\alpha_x:=\frac{S^*}{X^*}=\frac{G^*}{X^*}, \alpha_y = \frac{G^*}{Y^*}=\frac{M^*}{Y^*}.
\end{equation}  
In order to verify the equalities on the right-hand side of these equations substitute the normalized steady state $(x^*,y^*)=(1,1)$ into Eqs.~({\ref{eqPredPreyNorm}}), use $s(1)=g(1,1)=m(1)=1$, and demand for consistency $\dot{x}=\dot{y}=0$.

Using $\alpha_x$ and $\alpha_y$ we can write Eqs.~(\ref{eqPredPreyNorm}) as 
\begin{equation}
\label{eqPredPreyNormII}
\begin{array}{r c l}
\displaystyle \dot{x}&=&\alpha_x(s(x)-g(x,y))\\
\displaystyle \dot{y}&=&\alpha_y(g(x,y)-m(y))\\
\end{array}
\end{equation} 
In these equations $\alpha_x$ and $\alpha_y$ appear as parameters. Indeed they can be treated exactly like the parameters that are commonly used in conventional models. And, just like them they have an intuitive interpretation: From the definition, Eq.~(\ref{eqParamDef}), we can see that they denote the per capita gain and loss rates in the steady state. In other words, $\alpha_x$ is the turnover rate of prey, while $\alpha_y$ is the turnover rate of the predator. 

In order to investigate the local dynamics around the steady state we need to compute the system's Jacobian, which constitutes a local linearization of the system around the steady state under consideration. In case of our predator-prey system the Jacobian can be written as
\begin{equation}
\label{eqPredPreyJac}
{\rm \bf J}=\left( \begin{array}{c c} \alpha_x ( s_{\rm x} - g_{\rm x}) & -\alpha_x g_{\rm y} \\ 
  \alpha_y g_{\rm x} & \alpha_y (g_{\rm y} - m_{\rm y})\\\end{array}\right)  
\end{equation}    
where roman indices indicate derivatives of the respective function in the steady state. Specifically
\begin{eqnarray}
s_{\rm x}&=&\left. \partial s(x) /\partial x \right|_1, \\\nonumber
g_{\rm x}&=&\left. \partial g(x,y) /\partial x \right|_{(1,1)},\\\nonumber  
g_{\rm y}&=&\left. \partial g(x,y) /\partial y \right|_{(1,1)}, \\\nonumber
m_{\rm y}&=&\left. \partial m(y) /\partial y \right|_1.
\end{eqnarray}
Like the alphas $s_{\rm x}$, $g_{\rm x}$, $g_{\rm y}$ and $m_{\rm y}$ are constants for a given steady state and can therefore be considered as parameters. In order to understand how these parameters are interpreted let us consider how the normalization procedure would affect a specific function. Suppose one of the functions in the model, say $M(Y)$, was linear. 
In this case the corresponding normalized function were a linear function with slope 1, regardless of the slope  of $M(Y)$. Hence, for every linear $M(Y)$ the corresponding parameter would be $m_{\rm y}=1$. In fact, for an arbitrary monomial $M(Y)=AY^p$ the corresponding parameter is $m_{\rm y}=p$. For a general function $M(Y)$ we can interpret the corresponding parameter as a measure of sensitivity of the fluxes to the variables \cite{populations}. 

In ecology the sensitivity of top-predator mortality to top-predator abundance $m_{\rm y}$ is a well known quantity: the so-called exponent of closure. Likewise, the other parameters that appear in the predator-prey system have a well-defined ecological meaning. The parameter $s_{\rm x}$ denotes the sensitivity of the population growth of the prey with respect to prey abundance. If the prey is not limited by other external factors apart from predation the dependence should be linear ($s_{\rm x}=1$). If by contrast the prey population is strongly limited by external factors its growth rate should be insensitive to the abundance ($s_{\rm x}=0$).
Perhaps the most important parameter for the dynamics is $g_{\rm x}$, which denotes the dependence of predation on prey abundance. A Lotka-Volterra (i.e., mass action) model corresponds to $g_{\rm x}=1$. If predator saturation is taken into account then the predation rate is less sensitive to prey abundance, and we find $0 \leq g_{\rm x} \leq 1$ depending on the strength of the saturation. Finally, the parameter $g_{\rm y}$ describes the dependence of the predation rate on predator abundance. In most models this dependence is assumed to be linear ($g_{\rm y}=1$). 
   
Let us now return to the stability of steady states. The Jacobian governs the time evolution of the system close to the steady state. If the real parts of all eigenvalues of the Jacobian are negative then trajectories that start close to the steady state converge to it exponentially. In this case the equilibrium is said to be locally asymptotically stable. If parameters are changed the stability may be lost if the variation causes eigenvalues to acquire positive real parts. Since the Jacobian is a real matrix this can happen in either of two different ways: A single real eigenvalue crosses the imaginary axis in the origin of the complex plane, or two eigenvalues cross the imaginary axis as a complex conjugate pair. The first case corresponds to a bifurcation of saddle-node-type, such as fold, pitchfork, and transcritical bifurcations, which in general change the number and/or stability of steady states. The second case corresponds to a Hopf bifurcation, which, at least transiently, gives rise to oscillatory behavior.

From the Jacobian, Eq.~(\ref{eqPredPreyJac}), we find that in our model bifurcations of saddle-node-type 
occur if  
\begin{equation}
g_{\rm x}=s_{\rm x}-\frac{s_{\rm x}g_{\rm y}}{m_{\rm y}}. \\
\end{equation} 
A Hopf bifurcation occurs at
\begin{equation}
\label{eqHopfCond}
g_{\rm x}=s_{\rm x}-\frac{\alpha_y}{\alpha_x}(g_{\rm y}-m_{\rm y}) 
\end{equation}
if $(g_{\rm x}-s_{\rm x})m_{\rm y}+s_{\rm x}g_{\rm y}>0$. Note that only the ratio $r=\frac{\alpha_y}{\alpha_x}$ appears in the bifurcation conditions. This is reasonable as the absolute values of the turnover rates can be changed by means of a timescale renormalization. 

\begin{figure}[htb]
\centering
\includegraphics[width=2.5in]{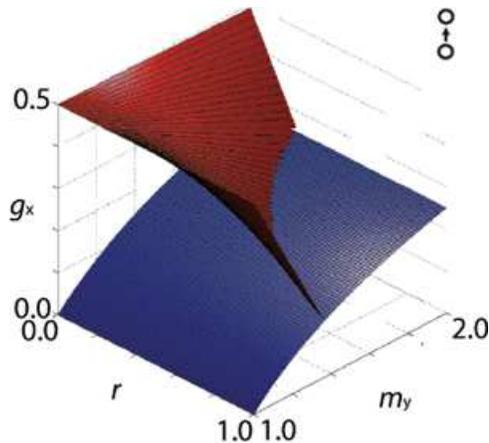}
\caption{Bifurcation diagram of a predator-prey system, depending on the sensitivity of the predator to prey abundance $g_{\rm x}$, the timescale separation $r$, and the exponent of closure $m_{\rm y}$. Steady states are stable in the top-most volume of parameter space. Stability is lost in a Hopf bifurcation (red surface) or in a bifurcation of saddle-node-type (blue surface). The sketch in the top-right corner indicates the topology of trophic interactions (Parameters: $s_{\rm x}=0.5$, $g_{\rm y}=1$.) \label{figPredPrey}}
\end{figure}

The critical parameter values at which the bifurcations occur form hyper-surfaces in parameter space. These surfaces are shown in Fig.~\ref{figPredPrey}. Every point in this diagram corresponds to steady states in 
a class of models with similar local dynamics. The steady states that fall into the topmost volume of parameter space are stable; all others are unstable. Destabilization can occur in a Hopf bifurcation (red) or in a bifurcation of saddle-node-type (blue). Among other things the diagram shows that a high sensitivity of predation to prey abundance (high $g_{\rm x}$) has a stabilizing effect on the system. 

Before we move on to the next section, let us recapitulate what we have achieved: Starting from a generalized model, Eq.~(\ref{eqPredPrey}), we have parameterized the Jacobian (and therefore the local dynamics) around every possible steady state in the class of models under consideration. Apart from the analytical computation of bifurcation points, which we have used here, the Jacobian can be analyzed in a number of ways: For instance we can use Monte-Carlo sampling to identify the parameters to which stability is most sensitive or we can use an optimization approach to find the most stable parameter set \cite{metabolic,TCA}. However, in this paper, we will show that certain topologically different food webs can yield exactly the same bifurcation diagrams, which is best done analytically.    
 
\subsection{A generalized model of competing predators}
Let us now consider a system in which two populations of predators feed on the same prey population. Using essentially the same notation as above, a generalized model for the system can be written as
\begin{equation}
\begin{array}{r c l}
\displaystyle \dot{X}&=& S(X)-G_1(X,Y_1)-G_2(X,Y_2)\\
\displaystyle \dot{Y_1} &=& G_1(X,Y_1) - M_1(Y_1)\\
\displaystyle \dot{Y_2} &=& G_2(X,Y_2) - M_2(Y_2)\end{array}
\end{equation}
The analytical treatment is very similar to the previous example, for this reason we skip some details of the normalization 
and jump directly to the equation corresponding to Eq.~(\ref{eqPredPreyNorm}), which now reads
\begin{equation}
\label{eqCompeteIntermediate}
\begin{array}{r c l}
\displaystyle \dot{x}&=& \frac{S^*}{X^*}s(x)-\frac{{G_1}^*}{X^*}g_1(x,y_1)-\frac{{G_2}^*}{X^*}g_2(x,y_2)\\
\displaystyle \dot{y_1} &=& \frac{{G_1}^*}{{Y_1}^*}g_1(x,y_1) - \frac{{M_1}^*}{{Y_1}^*}m_1(y_1)\\
\displaystyle \dot{y_2} &=& \frac{{G_2}^*}{{Y_2}^*}g_2(x,y_2) - \frac{{M_2}^*}{{Y_2}^*}m_2(y_2)
\end{array}
\end{equation}
In analogy to the predator prey system we can define
\begin{equation}
\alpha_{y1} = \frac{{G_1}^*}{{Y_1}^*}=\frac{{M_1}^*}{{Y_1}^*}, \alpha_{y2} = \frac{{G_2}^*}{{Y_2}^*}=\frac{{M_2}^*}{{Y_2}^*}
\end{equation} 
Note that in the first equation of Eq.~(\ref{eqCompeteIntermediate}) there are now three terms. We therefore, define
\begin{equation}
\alpha_x=\frac{S^*}{X^*}=\frac{{G_1}^*}{X^*}+\frac{{G_2}^*}{X^*}
\end{equation}
and 
\begin{equation}
\beta = \frac{{G_1}^*}{{G_1}^*+{G_2}^*}, \bar{\beta}=\frac{{G_2}^*}{{G_1}^*+{G_2}^*}
\end{equation}
which allows us to write Eq.~(\ref{eqCompeteIntermediate}) as 
\begin{equation}
\begin{array}{r c l}
\displaystyle \dot{x}&=& \alpha_x (s(X)-\beta g_1(x,y_1)- \bar{\beta} g_2(x,y_2))\\
\displaystyle \dot{y_1} &=& \alpha_{y1} (g_1(x,y_1) - m_1(y_1))\\
\displaystyle \dot{y_2} &=& \alpha_{y2} (g_2(x,y_2) - m_2(y_2))\\\end{array}
\end{equation}
In analogy to the predator-prey system we compute the Jacobian
\begin{equation}
{\rm \bf J}=\left( \begin{array}{c c c} \alpha_x ( s_{\rm x} - \beta g_{\rm1,x} - \bar{\beta} g_{\rm2,x}) & -\alpha_x \beta g_{\rm 1,y} & -\alpha_x \bar{\beta} g_{\rm 2,y}\\ 
  \alpha_{y1} g_{\rm 1,x} & \alpha_{y1} (g_{\rm 1,y} - m_{\rm 1,y}) & 0 \\
  \alpha_{y2} g_{\rm 2,x} & 0 & \alpha_{y2} (g_{\rm 2,y} - m_{\rm 2,y}) \\\end{array}\right)  
\end{equation}     
where now $a_{i,b}=\frac{\partial a_i}{\partial b_i}$. Note that a steady state in which all populations have positive size does generally exist in the system, as the canonical scenario of competitive-exclusion only occurs for $g_{\rm 1,y}=m_{\rm 1,y}$ and $g_{\rm 2, y}=m_{\rm 2,y}$ \cite{coexistence}.

The condition for bifurcations of saddle-node-type 
now reads 
\begin{equation}
\label{eqPreBifCont}
\begin{array}{c}
\displaystyle s_{\rm x}(g_{\rm 1,y}-m_{\rm 1,y})(g_{\rm 2,y}-m_{\rm 2,y})+\beta g_{\rm 1,x}m_{\rm 1,y}(g_{\rm 2,y}-m_{\rm 2,y})\\\displaystyle +\bar{\beta} g_{\rm 2,x}m_{\rm 2,y}(g_{\rm 1,y}-m_{\rm 1,y})=0 \end{array}
\end{equation}
Our aim is to compare this condition to the corresponding condition in the predator-prey system. This comparison is complicated by the fact that not only the state spaces but also the parameter spaces of the two models  differ in dimensionality. We therefore have to map the parameter space of the competition model to the lower dimensional parameter space of the predator-prey system. For simplicity we focus on a symmetrical situation in which both populations of predators behave identically. We set
\begin{equation}
\label{eqParamIdentity}
\begin{array}{c}
\displaystyle \beta=\bar{\beta}=0.5\\
\displaystyle \alpha_{y1}=\alpha_{y2}=\alpha_{y} \\
\displaystyle g_{\rm 1,x}=g_{\rm 2,x}=g_{\rm x} \\
\displaystyle g_{\rm 1,y}=g_{\rm 2,y}=g_{\rm y} \\
\displaystyle m_{\rm 1,y}=m_{\rm 2,y}=m_{\rm y} \\
\end{array}
\end{equation}   
While this situation is quite special, let us emphasize that it does not mean that we are talking about 
a single population of predators that we have split arbitrarily into two groups. Such a degenerate situation would, among other pathologies, lead to a rank deficient Jacobian. In truth, the two populations can be quite different. What Eq.~(\ref{eqParamIdentity}) implies is only that they have the same impact on the prey population and that they respond in the same way to perturbations. 

Using this assumption the bifurcation condition, Eq.~(\ref{eqPreBifCont}) simplifies to 
\begin{equation}
\label{eqPreBifContII}
\begin{array}{r c l}
0&=&s_{\rm x}(g_{\rm y}-m_{\rm y})(g_{\rm y}-m_{\rm y})+0.5 g_{\rm x}m_{\rm y}(g_{\rm y}-m_{\rm y})+0.5 g_{\rm x}m_{\rm y}(g_{\rm y}-m_{\rm y})\\
 &=&s_{\rm x}(g_{\rm y}-m_{\rm y})^2+ g_{\rm x}m_{\rm y}(g_{\rm y}-m_{\rm y}).\end{array}
\end{equation} 
Since $g_{\rm y}\neq m_{\rm y}$ this is equivalent to 
\begin{equation}
 0=s_{\rm x}(g_{\rm y}-m_{\rm y})+ g_{\rm x}m_{\rm y}
\end{equation} 
and therefore 
\begin{equation}
g_{\rm x}=s_{\rm x}-\frac{s_{\rm x}g_{\rm y}}{m_{\rm y}}.
\end{equation} 
This is exactly the condition for bifurcations of saddle-node-type that we found in the predator-system. Also, 
under the assumptions stated in Eq.~(\ref{eqParamIdentity}), the condition for Hopf bifurcations in the competition model maps exactly onto the corresponding condition, Eq.~(\ref{eqHopfCond}), in the predator-prey system. 

\begin{figure}[htb]
\centering
\includegraphics[width=2.5in]{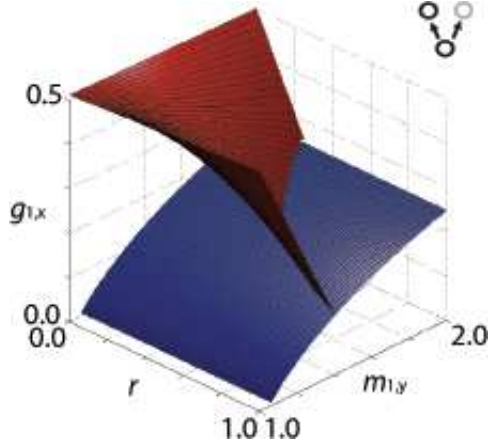}
\caption{Bifurcation diagram of a system of two predators competing for a common prey (see sketch in the top right corner). The generalized parameters describing predator 2 have been chosen to be slightly different ($\alpha_{y2}=1.05 \alpha_{y1}$, $m_{\rm 2,y}=0.95 m_{\rm 1,y})$ $g_{\rm 2,x}=0.95 g_{\rm 2,y}$.) Predator 2 causes 55 \% of the biomass loss of the prey.  As in Fig.~\ref{figPredPrey} the surfaces mark Hopf bifurcations points (red) and bifurcations points of saddle-node-type (blue). Although the diagram is not identical to Fig.~\ref{figPredPrey} it shows a strong similarity. (Further parameters: $s_{\rm x}=0.5$, $g_{\rm y}=1$.) \label{figModification}}
\end{figure}

Let us now see what happens if we relax Eq.~(\ref{eqParamIdentity}). For this purpose we change all parameters describing one of the competitors slightly. The resulting bifurcation diagram is shown in Fig.~\ref{figModification}. While the bifurcation surfaces shift a little, the structure of the bifurcation diagram is qualitatively conserved: No new bifurcation surfaces appear or vanish in response to the change of parameters. It is conceivable that the two diagrams in the figure could still be exactly identical if we had chosen the mapping of parameters on the axes of the diagram more cleverly. This last point will be discussed in more detail in Sec.~\ref{secGeneralized}. 

To summarize: We have used the approach of generalized modeling to identify a situation in which the local bifurcation diagrams of a predator-prey system is exactly identical to the corresponding diagram of a system describing exploitative competition. While this situation, is admittedly special in the context of the generalized model, it can be found in the whole family of conventional models that satisfy Eq.~(\ref{eqParamIdentity}). The equivalence of bifurcation diagrams implies that the dynamics of the two models is, at least locally, identical and therefore the coarse-grained (predator-prey) model and the full (exploitative competition) model have similar local dynamics. 
  
\begin{figure}[htb]
\centering
\includegraphics[width=5in,height=2in]{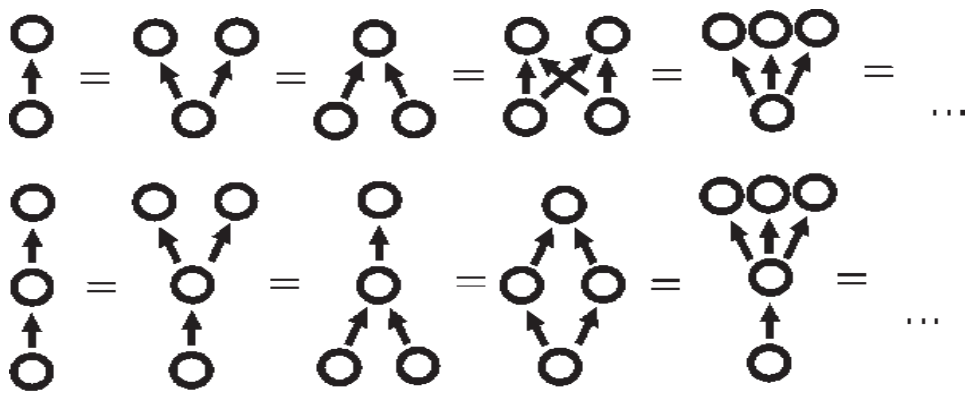}
\caption{Symbolical representation of some examples of the equivalence relation between bifurcation diagrams of different food webs. The sketches indicate food web topologies. If species on the same trophic level interact with the same neighbors and are described by the same generalized parameters they can be merged without altering the local bifurcation diagram. In the numerical verification of the relations shown here passive prey switching has been assumed. \label{figSymbolical}}
\end{figure}  
     
\section{Equivalence of more complex food webs \label{secGeneralized}}
In the previous section we have shown that under certain conditions different systems may be described 
by exactly the same local bifurcation diagrams. This insight is not limited to the minimal examples studied 
in the previous section but can be observed in many food webs. The Jacobian of a very general food web was derived in \cite{thesis} and in a slightly different parameterization in \cite{populations}. This Jacobian describes the local dynamics in complex food webs with nonlinear interactions and takes passive prey switching into account. We have compared the bifurcation diagrams of 43 different food webs numerically. Some of the results are shown symbolically in Fig.~\ref{figSymbolical}. 

Based on our results we conjecture the following: If two or more populations in the system interact with exactly the same neighbors and their generalized parameters are identical in the sense of Eq.~(\ref{eqParamIdentity}) then the populations can be modeled by a single dynamical equation with the corresponding parameters. Again we emphasize that identical generalized parameters do not imply an identity of organisms. 

In the following we call two food webs \emph{equivalent} if they satisfy the assumptions of the conjecture. In this case we say that the food web with the higher number of populations is \emph{reducible} to the food web with the lower number of populations. 
  
\begin{figure}[htb]
\centering
\includegraphics[width=2.5in]{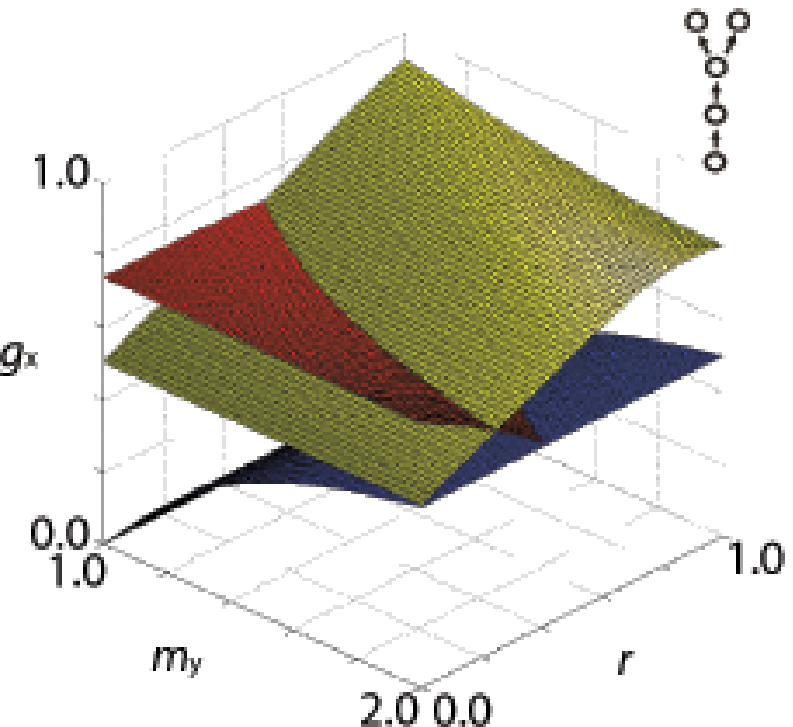}
\includegraphics[width=2.5in]{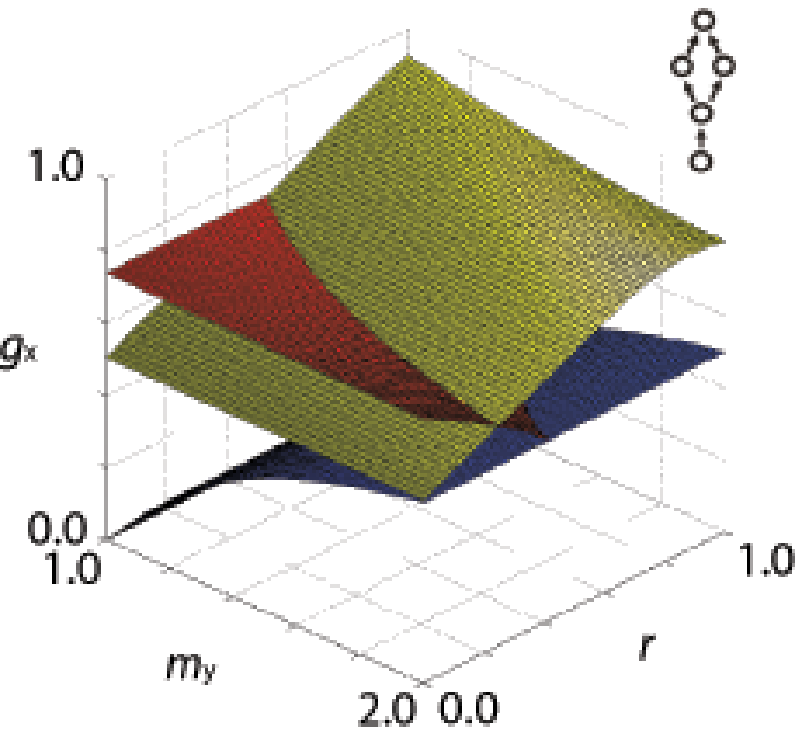}\\
\includegraphics[width=2.5in]{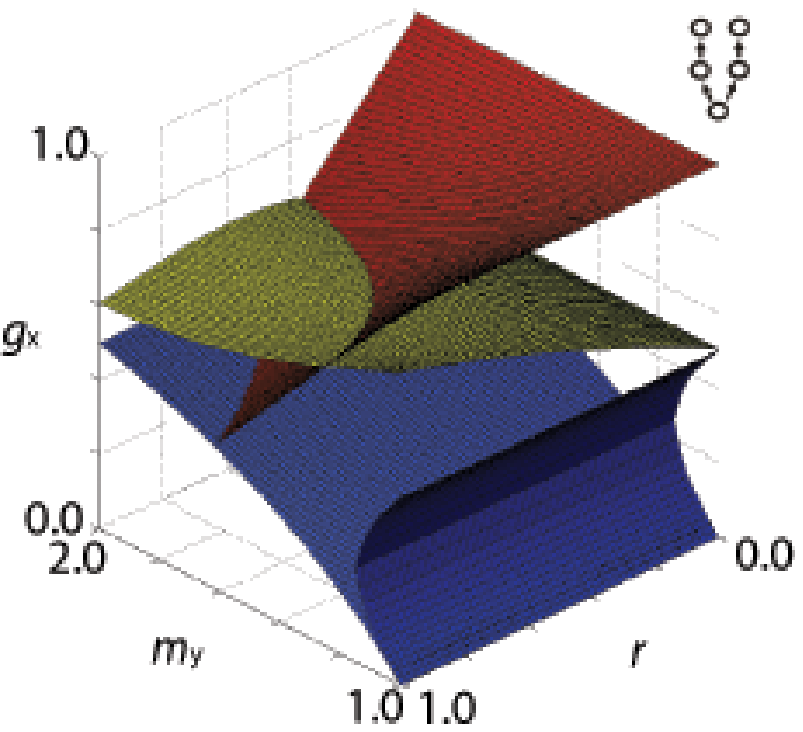}
\includegraphics[width=2.5in]{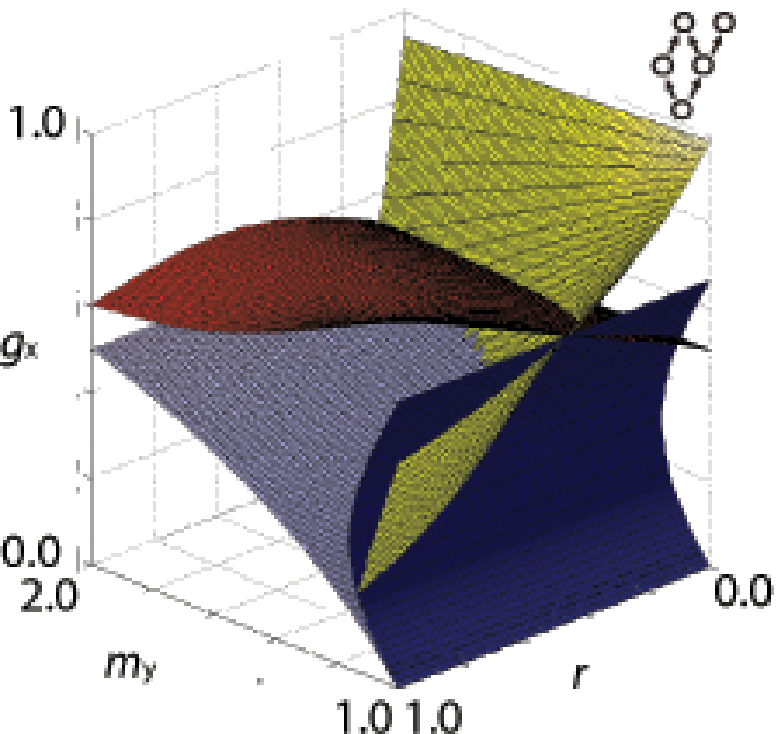}
\caption{Top row: Equivalence of complex food webs. If parameters are chosen appropriately (see text) the bifurcation diagrams of certain complex food webs are exactly identical. The respective topologies are indicated in the top right corner. Both diagrams are identical to the corresponding diagram of a four-trophic food chain.
Bottom row: Example of two food webs that cannot be mapped to food chains in the proposed way. 
In all four diagrams the red, yellow, and green surfaces correspond to Hopf bifurcations and the blue surface to bifurcations of saddle-node type  (Parameters: $s_{\rm x}=0.5$, $g_{i,{\rm y}}=1$, 
$g_{i{\rm x}}=g_{\rm x}$, $\alpha_{i}=r^j$ where $j$ is the trophic level of $i$, $m_{i,{\rm y}}=m_{\rm y}$. For details see \cite{populations}) \label{figReducibleFour}}
\end{figure}  
  
In particular the conjecture implies that there is a class of food webs that are reducible to food chains of the same number of trophic levels. An example of such food webs is shown in Fig.~\ref{figReducibleFour} (top panels). If one of the assumptions of the conjecture is violated we generally find that the food webs are not equivalent under a mapping of the form of Eq.~(\ref{eqParamIdentity}) (see bottom panels of  Fig.~\ref{figReducibleFour}). In particular a food chain can not be reduced to another food chain with fewer trophic levels.

Additionally, we observed that the bifurcation diagram of a given food web contains at least as many Hopf bifurcation surfaces as the food chain with the same number of trophic levels, while food webs that are not 
reducible to food chains generally have additional bifurcation surfaces. Two examples of such food webs that are not reducible to food chains are also shown in Fig.~\ref{figReducibleFour} (bottom panels). 

The examples discussed above show that the food web topology has a strong impact on reducibility. 
We therefore have to ask what happens if we alter the topology of a reducible food web. 
At which point will the transition to irreducibility occur? In our general food web model the topology 
is captured by a number of parameters, that encode the strength of topological connections in analogy
to the parameter $\beta$ in the competition model from the previous section. We can tune these parameters continuously between reducible and irreducible topologies.

\begin{figure}[htb]
\centering
\includegraphics[width=2.5in]{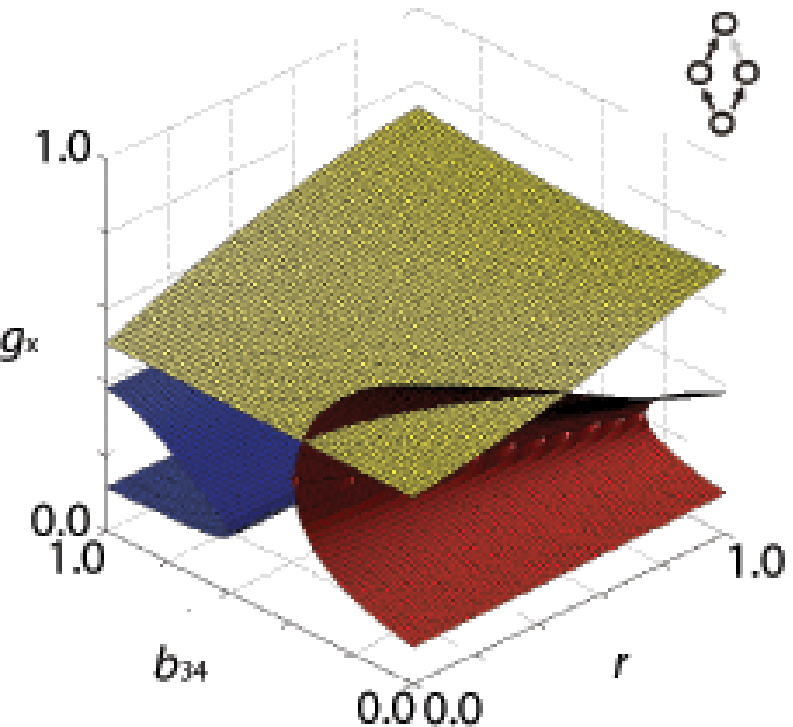}
\includegraphics[width=2.5in]{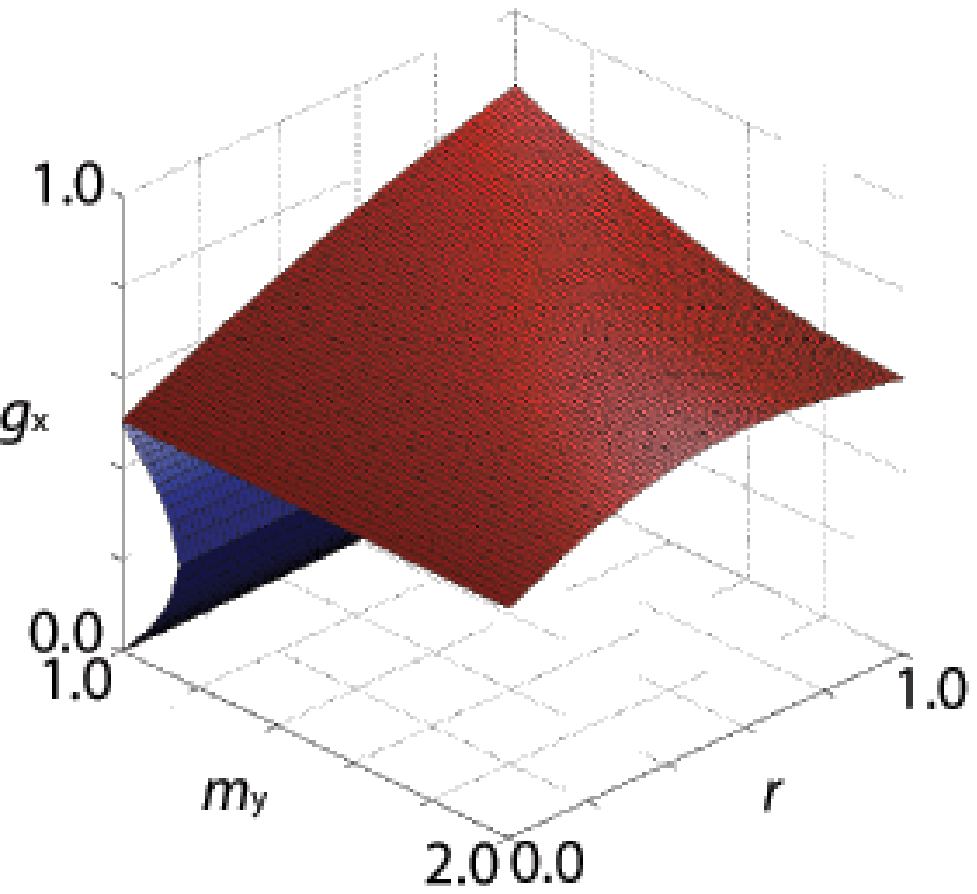}\\
\includegraphics[width=1in]{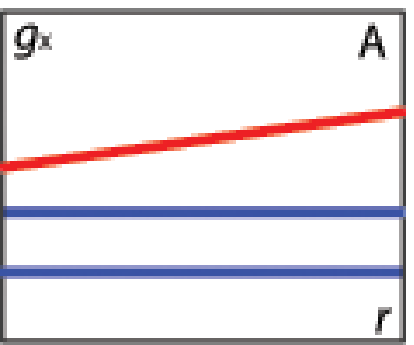}
\includegraphics[width=1in]{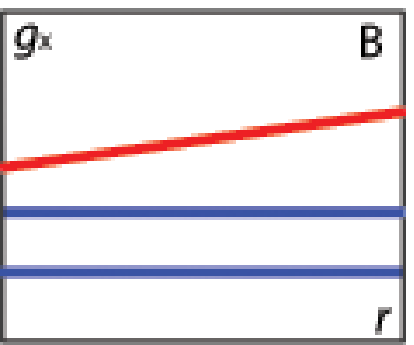}
\includegraphics[width=1in]{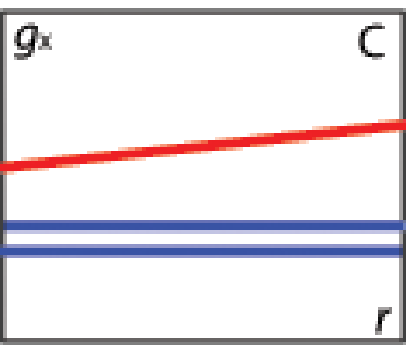}
\includegraphics[width=1in]{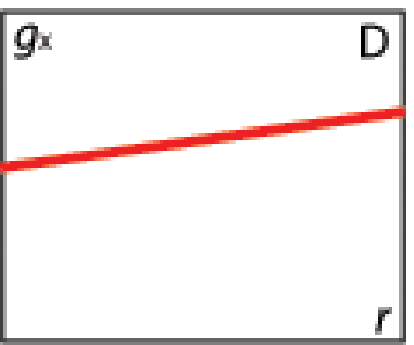}
\includegraphics[width=1in]{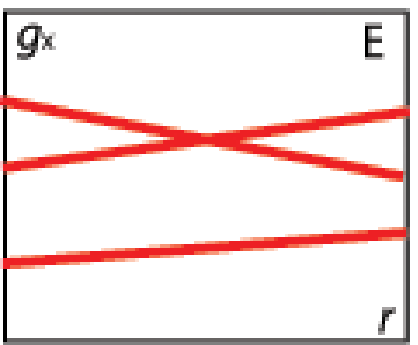}
\caption{Top row: Comparison of the bifurcation diagram of a tri-trophic food chain (top right) with a food web of variable topology (top left). In the variable food web the strength of one trophic link is controlled by the parameter $b_{3,4}$. This link is shown in grey in the symbolic representation of the food web topology. For parameters see Fig.~\ref{figReducibleFour} except for $m_{\rm y}$, which is 1.1 in the left diagram. 
Bottom row: sketches of two-parameter slices through the three-parameter diagrams. The two parameter diagram for the tri-trophic food chain at $m_{\rm y}=1.1$ (A) is exactly identical to the diagram of the variable system for $b_{3,4}=1$ (B). If $b_{3,4}$ is decreased the bifurcation diagram changes (C) but remains qualitatively similar to A. At still lower values (D) the bifurcation diagram differs qualitatively from the diagram of the food chain for $m_{\rm y}=1.1$ but is still similar to the food chain diagram at higher $m_{\rm y}$. Finally, at low values of $b_{3,4}$ the bifurcation diagram of the variable system is no longer similar to the bifurcation diagram of the tri-trophic food chain for any set of parameter values (E). \label{figWebBreak}}
\end{figure}

Let us illustrate this point in a simple example. We consider a system in which two competing populations of predators feed on a single prey population. Both predator populations are consumed by a population of top predators. In our model the prey species is population $1$, the competing predators are population $2$ and $3$ and the top-predator is population $4$ (see sketch in Fig.~\ref{figWebBreak}). 
We assume that both populations of predators are described by similar generalized parameters in the sense 
of Eq.~(\ref{eqParamIdentity}), except that the biomass flow from population 3 to population 4 may be less than the biomass flow from population 2 to population 4. We can imagine that population 3 has some defense against 
the predator that population 2 lacks. We measure the efficiency of this defense, by a parameter $b_{3,4}$, 
which indicates the strength of the trophic link. If $b_{3,4}=1$ the defense is ineffective, and individuals 
of population 3 are as likely to be consumed by the top predator as individuals of population 2. We assume that in this case natural mortality of individuals of population 2 and 3 can be neglected. If we decrease $b_{3,4}$ the efficiency of the defense increases. In this case only a certain portion of the biomass loss of population 3 is caused by the top predator while the remainder is due to natural mortality. If $b_{3,4}=0$ then the biomass loss of population 3 because of predation vanishes, and the entire biomass loss of population 3 
occurs because of natural mortality. 


A bifurcation diagram of the system described above is shown in Fig.~\ref{figWebBreak}. In this diagram 
we have chosen $b_{3,4}$ as one of the axes. 
For $b_{3,4}=0$ the food web can be reduced to the tri-trophic food chain. Therefore, the parameter plane with $b_{3,4}=0$ is exactly identical to a two-parameter bifurcation diagram of this food chain at the corresponding exponent of closure $m_{\rm y}=1.1$. 
As we decrease $b_{3,4}$ the exact identity is lost but the topology of the bifurcation surfaces at first stays qualitatively similar. 
At a certain value of $b_{3,4}$ the bifurcation surface of saddle-node-type folds back and disappears. 
From this point the two-parameter bifurcation diagram that we find in the food web for a given value of $b_{3,4}$ is no longer equivalent to the bifurcation diagram of the tri-trophic food chain at the corresponding value of $m_{\rm y}$. 
However, note that the two-parameter diagram is still qualitatively similar to a two-parameter diagram of the food chain at a higher exponent of closure $m_{\rm y}$. 
If $b_{3,4}$ is decreased further then a new Hopf bifurcation surface appears. 
The bifurcation diagram is now qualitatively different from the bifurcation diagram of the food chain. In particular, double Hopf-bifurcations, i.e.~intersections of Hopf bifurcation surfaces as shown in Fig.~\ref{figWebBreak} (E), can not appear in our tri-trophic food chain model at all. 

Above we have conjectured that local bifurcation diagrams of a class of food webs are exactly reducible to the corresponding diagrams of food chains with the same number of trophic levels. Figure \ref{figWebBreak} shows that in the example a larger class of food web appears to exhibit qualitatively similar local dynamics. These  insights, if proved, can be used to extend certain results on food chains to classes of food webs.

One question that has been discussed in the ecological literature for quite some time is whether chaotic dynamics occur in natural food webs \cite{May:ChaosReview,Upadhyay:ChaosReview,Rai:ChaosReview}. On the one hand even some of the simplest ecological models exhibit chaotic dynamics (e.g.~\cite{May:Chaos,May:LogisticMap,Edwards:SteeleModel}). On the other hand chaos has rarely been observed in nature \cite{Cushing:PopDynReview,Tilman:Grasses,Hanski:Rodents,Turchin:Voles}. In the past it has been argued that chaotic dynamics could be detrimental for the survival of the participating populations and could therefore disappear in the course of natural selection. However, selection acts primarily on the level of the individual. And, even on the level of the population there is some evidence that chaos, if considered in a spatial or meta-population context, promotes persistence \cite{Allen:Spatial,Sole:Spatial,Petrovskii:Spatial}. Perhaps the most widely accepted view is now that chaos may be present in the dynamics but is difficult to detect because of environmental noise \cite{Nychka:Noise,Ellner:Noise}. 
However, another opinion is that the complexity of large ecological systems somehow prevents chaotic dynamics \cite{Steele:Predation,Ruxton:PopulationFloors,Fussmann:Chaos}.

Since the approach of generalized modeling focuses on local dynamics, global dynamics such as chaos is not directly accessible. 
However, we can draw conclusions on certain features of global dynamics by considering bifurcations of higher codimension. 
Unlike Hopf and saddle-node bifurcations, which are of codimension 1, bifurcations of higher codimension have more than one bifurcation condition. In comparison to codimension-1 bifurcations the higher codimension bifurcations are often neglected, since the tuning of more than one parameter is necessary to find them in experiments or numerical investigations. 

An example of a codimension-2 bifurcation is the double-Hopf bifurcation, formed at the intersection of two Hopf bifurcation surfaces, which we have already encountered above. Mathematical investigations of this bifurcation have shown that there is generically a region in parameter space in which chaotic dynamics occur close to the double-Hopf bifurcation point \cite{Kuznetsov:Elements}. In this parameter region chaotic dynamics can therefore, at least transiently, be observed.

In a previous paper we have shown that double Hopf bifurcations exist in food chains with more than three trophic levels. This implies that chaotic dynamics can be found generically in these food chains at least in some parameter space. 
Our conjecture implies that a certain class of food webs is exactly equivalent to food chains of the same length. 
This means that these food webs also exhibit double-Hopf bifurcations and therefore chaotic parameter regions. Furthermore, double-Hopf bifurcations should also exist in the much larger class of food webs in which the bifurcation diagrams are qualitatively similar to the bifurcation diagrams of food chains. 
Finally, irreducible food webs in which the local dynamics differs qualitatively from that of food chains have, in our experience, a more complex bifurcation structure than food chains of comparable length and therefore the corresponding bifurcation diagrams also contain double-Hopf bifurcations. 

From the evidence given above we conclude that chaotic parameter regions should exist in a large class of complex food webs with more than three trophic levels. It is therefore unlikely that the complexity of natural food webs in itself leads to the avoidance of chaotic dynamics.        

\section{Remarks and Discussion \label{secConclusions}}
\label{secRemarks} 
In this paper we have used the approach of generalized models to show that in certain food webs the local bifurcation diagrams are exactly identical to the corresponding bifurcation diagrams in other 
food webs and food chains. 
Moreover, we have conjectured that there is a universal rule that governs which food webs show this equivalence. This conjecture, if proved, can be used to generalize certain findings on food chains to classes of food webs. 
Perhaps more importantly, it holds the promise to yield an analytical procedure for the reduction of complex food-web models or even dynamical biological networks in general. 

A method for coarse-graining food webs based on the equivalence of local bifurcations had several advantages: It would probably preserve local dynamics exactly, and also preserve certain features of the global dynamics through conservation of local bifurcations of higher codimension.  
Moreover, it would probably provide hard criteria governing which populations could be coarse-grained into a single variable and which could not. 

Before the present observations can be extended to a method for model reduction more research is certainly necessary.  
This work should aim to understand the equivalence, prove the stated conjecture and in particular explore its validity in the large class of systems in which we so far only observe a qualitative equivalence of bifurcation diagrams. Let us therefore conclude with some remarks that we believe will be conductive for this effort. 

One feature of the mapping of a food web to a simpler food web is that it reduces the dimensionality of both the state space and the parameter space. In the context of our approach the mapping in state space is not of much concern as we do not need to specify it explicitly. By contrast the mapping in parameter space is more intriguing; We have to specify how the parameters of the original model are translated to the lower number of parameters in the reduced model. 

In this paper we have postulated the mapping of parameters \emph{ad hoc} based on biological reasoning and only subsequently verified that it produced the desired result. For instance in our discussion of the system of competing predators we have demanded that both predators exhibit the same exponent of closure $m_{\rm y}=m_{\rm 1,y}=m_{\rm 2,y}$ (Eq.~\ref{eqParamIdentity}). Only thereafter we could compare the bifurcation diagram to that of the predator-prey system. For future investigations a different approach seems more promising: Instead of demanding the identity of certain parameters and then checking for the identity of bifurcation diagrams, we could ask which mapping of parameters, say $m_{\rm y}(m_{\rm 1,y},m_{\rm 2,y})$ will produce a bifurcation diagram that is identical to the one of the simpler system. Of course we do not need to restrict this mapping to parameters between which a direct biological correspondence exists, but could allow the parameter of the reduced system $m_{\rm y}$ also to depend on other parameters such as $g_{\rm 1,y}$ and $g_{\rm 2,y}$. While more mathematical work is certainly necessary, we believe that it should be possible to derive a rigorous procedure for the parameter mapping.

We believe that the procedure described above will reveal exact reductions for the large class of systems in which we so far only observe a qualitative similarity of bifurcation diagrams. Nevertheless, it is obvious that for certain pairs of food webs there cannot be any mapping of parameters that leads to identical bifurcation diagrams. For instance, we have seen that the bifurcation diagram of the four-trophic food chain contains a double-Hopf bifurcation line. Since this bifurcation is characterized by four purely imaginary eigenvalues of the Jacobian it simply can not be accommodated in any system with less than four state variables. Therefore the local bifurcation diagram of the four-trophic food chain can never be mapped on that of the tri-trophic food chain.    

The fact that generalized modeling reveals bifurcations of higher codimension with relative ease is advantageous as it provides examples of these bifurcations for mathematical analysis and allows us to gain insights into global features of the dynamics based on a local analysis. In the present paper this has be utilized to provide evidence for chaos in a large class of food webs. This and other bifurcations also reveal the presense of global homoclinic and heteroclinic bifurcations. It is conceivable that future progress in normal form theory will enable us to extract even more information from the bifurcations of higher codimension.  

So far we have focused on generalized models while in the literature most modeling is done by conventional models in which the interactions are restricted to specific functional forms. Note however, 
that it is straight forward to step back and forth between generalized and conventional representations of a system. By replacing the specific functional forms by general functions, a conventional model can be turned into a generalized model and vice versa \cite{chaos}. The corresponding mapping between generalized and specific model parameters can be computed analytically and thus provides a direct relation between the modeling approaches \cite{george}. For a smoother transition, it is possible to study hybrid models in which some functional relationships are specified while others are kept general \cite{populations}.  

Let us emphasize that the identity of bifurcation diagrams is therefore not restricted to the realm of generalized models; One should be able to observe the same identity also in conventional models. 
However, in conventional models additional difficulties arise since the mapping is affected by the 
position of steady states in state space which is often prohibitively difficult to study analytically. 
In the past we have pointed out that the parameters that are identified in generalized modeling are in a certain sense more natural than the parameters that are commonly introduced in conventional modeling \cite{thesis,populations}. In the present paper this enabled us to spot the identity of bifurcation diagrams even based on very simple ad hoc assumptions. While the same identity should exist also in conventional models the corresponding mapping of the conventional parameters is probably much more involved and hence the identity is more difficult to spot.

To conclude: We have shown that in generalized models an exact equivalence between the bifurcation diagrams of different food webs can be observed, albeit in a relatively special case. Several results seem to indicate that this identity can be extended to a much larger class of systems if more refined parameter mappings are used. We hope that this insight will eventually evolve into a method for model reduction. In the context of such a method an intermediate coarse-graining step with generalized models will probably be useful even if the desired result and starting point are conventional models.  

\section*{Acknowlements}
This work was supported by the Deutsche Forschungsgemeinschaft (DFG), grant FE 359.

\bibliographystyle{elsart-num}
\bibliography{Coexistence,DynamicsBooks,EcologyBooks,MyPapers,Plankton,EcoChaos,NetworkAnalysis}
\end{document}